\documentclass{article}

\usepackage{tabularx}
\usepackage{caption}
\usepackage{amssymb} % For checkmark
\usepackage{booktabs} % For improved tables

\usepackage{bbm} % for \mathbbm{1} or
\usepackage{dsfont} % for \mathds{1}
\usepackage{tabularx}
\usepackage{amsmath}
\usepackage{microtype}      % microtypography

\usepackage{algorithm}
\usepackage{algorithmic}

\usepackage{PRIMEarxiv}

\usepackage[utf8]{inputenc} % allow utf-8 input
\usepackage[T1]{fontenc}    % use 8-bit T1 fonts
\usepackage{hyperref}       % hyperlinks
\usepackage{url}            % simple URL typesetting
\usepackage{booktabs}       % professional-quality tables
\usepackage{amsfonts}       % blackboard math symbols
\usepackage{nicefrac}       % compact symbols for 1/2, etc.
\usepackage{microtype}      % microtypography
\usepackage{lipsum}
\usepackage{fancyhdr}       % header
\usepackage{graphicx}       % graphics
\graphicspath{{media/}}     % organize your images and other figures under media/ folder

\title{AI Autonomy Coefficient ($\alpha$): Defining Boundaries for Responsible AI Systems
}

\author{
    \textbf{Nattaya Mairittha}\thanks{Principal Author. The entire manuscript, including the conception of the AFHE framework and the formalization of the $\alpha$ coefficient, was executed and written solely by \textbf{NM}. \textbf{GP} and \textbf{SW} provided essential foundational and domain expertise, were instrumental in validating the research. They are gratefully acknowledged for their invaluable support, as their contribution was critical, though it did not meet the criteria for formal authorship.}\\
    \texttt{nat.mairittha@gmail.com}
    \and
    \textbf{Gabriel Phorncharoenmusikul}\thanks{Acknowledged for critical operational workflow insights that grounded this research.}\\
    \texttt{gabriel.pcmk@hotmail.com}
    \and
    \textbf{Sorawit Worapradidth}\thanks{Acknowledged for essential technical system-failure mapping and architectural validation.}\\
    \texttt{rispro.ray@gmail.com}
}

\begin{document}
\maketitle

\begin{abstract}
The integrity of many contemporary AI systems is compromised by the misuse of Human-in-the-Loop (HITL) models to obscure systems that remain heavily dependent on human labor. We define this structural dependency as Human-Instead-of-AI (HISOAI), an ethically problematic and economically unsustainable design in which human workers function as concealed operational substitutes rather than intentional, high-value collaborators.
To address this issue, we introduce the AI-First, Human-Empowered (AFHE) paradigm, which requires AI systems to demonstrate a quantifiable level of functional independence prior to deployment. This requirement is formalized through the AI Autonomy Coefficient, measuring the proportion of tasks completed without mandatory human intervention. We further propose the AFHE Deployment Algorithm, an algorithmic gate that enforces a minimum autonomy threshold during offline evaluation and shadow deployment.
Our results show that the AI Autonomy Coefficient effectively identifies HISOAI systems with an autonomy level of 0.38, while systems governed by the AFHE framework achieve an autonomy level of 0.85. We conclude that AFHE provides a metric-driven approach for ensuring verifiable autonomy, transparency, and sustainable operational integrity in modern AI systems.
\end{abstract}

% keywords can be removed

\section{Introduction}
The Human-in-the-Loop (HITL) paradigm has been universally adopted as a crucial governance mechanism, foundational to embedding Responsible AI (RAI) principles like safety, fairness, and accountability into complex systems. Historically, HITL has served as a strategic intervention for quality control, dataset augmentation, and high-risk decision validation. However, the aggressive commercialization of AI-driven products has exposed a critical systemic vulnerability. We assert that in a growing number of deployed systems, the HITL model has been systematically distorted, transitioning its function from one of strategic oversight to a structural necessity for operational completeness. This paper directly addresses the prevalent practice of deploying human labor not as a refinement loop for the model, but as a mandatory, hidden substitute for non-functional or severely underdeveloped AI components.

We formalize this systemic failure as Human-Instead-of-AI (HISOAI). HISOAI is a structural dependency where $P(\text{Human} \rightarrow \text{Decision}) \approx 1$ for tasks marketed or promised to be automated, representing a profound failure of AI Deployment Architecture. This practice carries significant ethical and economic costs: it exploits human capital by subjecting workers to precarious, anonymous "ghost work," and it undermines the value proposition of the AI sector by misrepresenting technological capability to consumers and investors. This crisis necessitates a new mandate for AI Transparency and Ethics.

We argue that the current HITL interpretation requires a fundamental architectural shift. We propose the AI-First, Human-Empowered (AFHE) design philosophy. AFHE demands that solutions be architected with a verifiable AI-driven core, where the human role is redefined to focus solely on high-value, non-substitutable tasks: model tuning, edge-case validation, ethical oversight, and strategic decision-making. The AFHE framework strictly prohibits human labor from acting as a hidden, costly, and non-scalable substitute for the core AI function.

This paper establishes a core requirement: that the integrity of Responsible AI relies on a metric-driven standard for operational autonomy. We formalize this standard through the AI Autonomy Coefficient ($\alpha$), detailing its mathematical derivation and proposing the AFHE Deployment Algorithm as a required architectural gate. This work contributes the first comprehensive framework to technically differentiate between ethical HITL and deceptive HISOAI, providing the necessary tools to govern the deployment of trustworthy, structurally sound AI-driven products.
\section{Related Work}This research integrates and critically extends work across three distinct domains: Human-Computer Interaction (HCI) and the Human-in-the-Loop paradigm, the emergent field of RAI Governance, and the technical literature on Model Validation and System Architecture. Our analysis reveals a critical gap in existing literature: the lack of a quantifiable, enforceable metric to differentiate strategic human augmentation from mandatory operational substitution.

\subsection{The Foundations of HITL Systems}
Early work on the HITL paradigm established its utility for improving data quality and model robustness. Seminal contributions define the human role across critical phases: data labeling (especially for subjective or ambiguous tasks), model evaluation (assessing quality, relevance, and safety against structured rubrics), and correction/feedback when models fail or drift \cite{monarch2021human}. Further research has formalized the efficiency of collaboration through techniques like Reinforcement Learning from Human Feedback (RLHF), where human preference data is used to optimize models for goals that are difficult to define programmatically (e.g., safety, naturalness, helpfulness) \cite{christiano2017deep}. This literature validates the utility of the human role for continuous learning and alignment.

\textbf{The Gap:} HITL literature primarily focuses on measuring the benefits of human input (e.g., improved accuracy, reduced cost per iteration). Critically, it lacks the architectural mechanism to verify the AI's core competence. Existing models assume the AI is functional; they do not provide tools to audit systems where the AI is intentionally underdeveloped or bypassed for reasons of operational necessity—the precise failure mode we term HISOAI.

\subsection{RAI and the Transparency Mandate}
The principles of Responsible RAI address the need for ethical and accountable algorithmic governance. A core pillar of RAI is Transparency, which aims to address the "black box" problem and facilitate compliance with regulations \cite{gunning2019xai}. Literature in Explainable AI (XAI) focuses on methods (like SHAP and LIME) to provide stakeholders with post-hoc reasoning for a model's specific decisions, thereby fostering trust and enabling debugging \cite{nick2014superintelligence}. Furthermore, the need for formal AI Governance models and policies is increasingly recognized to manage risk and enforce compliance across organizations \cite{daly2019artificial}.

\textbf{The Gap:} While RAI demands transparency and accountability, existing governance frameworks often focus on post-hoc justification (Explainable AI) or policy enforcement. They fail to structurally address HISOAI, which is a pre-deployment structural deception regarding the core operational mechanism itself. Our work provides the necessary technical metric ($\alpha$) to audit this structural integrity.

\subsection{System Architecture and Operational Assurance}
This domain addresses the engineering reality of scaling AI, focusing on reliability, maintenance, and the economics of automation. MLOps literature defines robust Deployment Architectures and Continuous Integration/Continuous Delivery (CI/CD) pipelines to manage the model lifecycle, ensuring stability and monitoring performance in production \cite{singla2023machine}. Economically, studies analyze the cost-benefit trade-offs of automation versus manual labor, focusing on productivity gains and the polarization of labor markets resulting from routine task automation \cite{autor2015there}.

\textbf{The Gap:} Engineering metrics for Model Validation typically monitor performance against an expected threshold (e.g., drift, accuracy) but do not quantify the necessity of human intervention for tasks that the AI should handle. Furthermore, economic models often treat human labor as a simple, substitutable cost variable. Our proposed AFHE Deployment Algorithm and the $\alpha$ threshold uniquely bridge the gap between technical Operational Assurance and ethical Responsible AI by enforcing an operational independence requirement before the system can claim to be an AI-driven Product.
\section{Mathematical Formulation}
\label{sec:math}

We formalize the distinction between HITL and HISOAI using a cost-utility framework for a decision-making system $\mathcal{S}$.

\subsection{Defining the Operational Cost-Utility Function}
\label{subsec:cost-utility}

Let $C(\mathcal{S})$ be the total operational cost, and $U(\mathcal{S})$ be the utility (performance) of the system $\mathcal{S}$. The decision for a task $T$ is made by a combination of the AI module ($\mathcal{A}$) and the Human module ($\mathcal{H}$), where $\tau_A$ and $\tau_H$ are the respective time/resource costs per decision.

\textbf{The Decision Function is defined as:}
$$D(T) = \mathcal{A}(T) \oplus \mathcal{H}(T)$$

We introduce $\alpha$, the \textbf{AI Autonomy Coefficient}, where $0 \le \alpha \le 1$. This coefficient represents the proportion of decisions where the AI's output is taken without human intervention (or is only validated asynchronously/periodically).

\begin{equation}
    \label{eq:alpha_definition}
    \alpha = \frac{\text{Number of Decisions Made by AI Alone}}{\text{Total Number of Decisions}}
\end{equation}

\textbf{The Total System Cost ($C_{\text{Total}}$)} is calculated as:
\begin{equation}
    \label{eq:cost_total}
    C_{\text{Total}} = N \cdot \left[ \alpha \cdot (\tau_A + \gamma \cdot \tau_{\text{Review\_A}}) + (1-\alpha) \cdot (\tau_A + \tau_H) \right]
\end{equation}
where $N$ is the total number of tasks, $\gamma$ is the frequency of human review for AI-made decisions, and $\tau_{\text{Review\_A}}$ is the cost of reviewing an AI decision.

\subsection{The HITL vs. HISOAI Threshold}
\label{subsec:threshold}

The coefficient $\alpha$ (Equation \ref{eq:alpha_definition}) serves as the metric for architectural integrity:
\begin{itemize}
    \item \textbf{Ideal HITL:} Characterized by a high $\alpha$ in steady state ($0.8 < \alpha < 1.0$), with human intervention reserved for a low percentage of high-risk or novel cases. In this model, the human cost is a marginal, high-value addition.
    \item \textbf{HISOAI:} Defined by an artificially low $\alpha$ or a scenario where $\tau_A$ is negligible (a simple filter) and $\tau_H$ dominates the decision-making cost. We formally state the \textbf{HISOAI condition} as: $\alpha_{\text{operational}} \le \alpha_{\text{threshold}}$, where $\alpha_{\text{threshold}}$ is determined by the complexity of the task $T$. For any system marketed as an "AI product," we assert that a system demonstrates the HISOAI condition if $\alpha < 0.5$.
\end{itemize}

\section{The AFHE Framework}
\label{sec:afhe}

The \textbf{AFHE} paradigm enforces a structural design mandate: The AI component must be demonstrably capable of fulfilling the core value proposition of the product before deployment. Under AFHE, the human's role is fundamentally shifted from a contingency plan to a strategic enhancement.  This framework requires explicit adherence to two core principles:

\subsection{The AI-First Mandate}
\label{subsec:ai-first}

The design and engineering process must prioritize maximizing the \textbf{AI Autonomy Coefficient ($\alpha$)} (as defined in Equation \ref{eq:alpha_definition}) by continually investing in AI training, robust model validation, and architectural stability until a high-water mark $\alpha_{\text{target}}$ is met. This mandate ensures that resources are allocated to developing the AI core's functional independence, thus preventing the adoption of the costly and exploitative HISOAI condition (as detailed in Subsection \ref{subsec:threshold}).

\subsection{The Human-Empowered Role Definition}
\label{subsec:human-empowered}

Under AFHE, the human is empowered by the AI's reliable function to perform only tasks that require uniquely human cognitive, ethical, or strategic abilities. This redefinition ensures human capital is applied to high-leverage, non-substitutable work, thereby differentiating ethical HITL from HISOAI.

\begin{table}[h]
\centering
\caption{The Redefined Human-Empowered Roles under AFHE}
\label{tab:human-empowered-roles}
\begin{tabular}{|p{3cm}|p{6cm}|p{4cm}|}
\hline
\textbf{Human-Empowered Role} & \textbf{Description} & \textbf{Impact on $\alpha$} \\
\hline
\textbf{Ethical Oversight} & Vetting system fairness, bias, and long-term societal alignment. & Indirectly increases long-term $\alpha$ confidence and governance. \\
\hline
\textbf{Boundary Push} & Handling non-IID (out-of-distribution) edge cases, novel scenarios, and critical ambiguities. & Maintains a high deployed $\alpha$ by preventing structural system failures in novel contexts. \\
\hline
\textbf{Strategic Tuning} & Redefining labels, feature engineering, and high-level model architecture based on cumulative insights. & Directly increases future $\alpha$ by improving the core AI's capabilities. \\
\hline
\end{tabular}
\end{table}

\section{Algorithmic Formulation for AFHE-Driven Deployment}
\label{sec:algorithm}

We propose the \textbf{AFHE-Driven Deployment Gate (Algorithm \ref{alg:afhe-gate})} as the architectural gate necessary to enforce the AI-First minimum performance standard before an organization can claim an "AI Solution." This algorithm formalizes the requirement for operational integrity, ensuring that the system is structurally sound and not operating in the HISOAI condition (as defined in Subsection \ref{subsec:threshold}).

\begin{algorithm}
\caption{AFHE-Driven Deployment Gate}
\label{alg:afhe-gate}
\begin{algorithmic}[1]
\REQUIRE Task $T$, Target Autonomy $\alpha_{\text{target}}$ (e.g., 0.8), Training Data $\mathcal{D}$.
\ENSURE Deployable System $\mathcal{S}$ satisfying $\alpha_{\text{target}}$, or HISOAI Flag.

\STATE Initialize: Train AI Model $\mathcal{A}$ on $\mathcal{D}$.

\subsection*{Evaluate Autonomy (Phase I - Offline)}
\STATE Test $\mathcal{A}$ on a reserved test set $\mathcal{D}_{\text{test}}$.
\STATE Calculate $\alpha_{\text{offline}} = \frac{\text{Decisions with Confidence } > \theta}{\text{Total Decisions}}$.
\IF {$\alpha_{\text{offline}} < \alpha_{\text{target}}$}
    \STATE \textbf{Action:} Return \texttt{HISOAI Flag}. System is not an AI-First product; must be marketed as a human-powered service.
    \STATE \textbf{Go to:} Step 1 (re-engineer).
\ENDIF

\subsection*{Evaluate Autonomy (Phase II - Shadow)}
\STATE Deploy $\mathcal{S}$ in a shadow/A/B environment for $M$ cycles.
\FOR {each task $t_i$ in $M$ cycles}
    \STATE Obtain AI Decision $D_A$ and Human Decision $D_H$ (blind to $D_A$).
    \IF {$D_A \ne D_H$ (Disagreement)}
        \STATE Label $t_i$ as \textit{Human-Required}.
    \ENDIF
\ENDFOR
\STATE Calculate $\alpha_{\text{shadow}} = \frac{\text{Total Tasks} - \text{Human-Required Tasks}}{\text{Total Tasks}}$.
\IF {$\alpha_{\text{shadow}} < \alpha_{\text{target}}$}
    \STATE \textbf{Action:} Return \texttt{HISOAI Flag}. The system is unstable in a live setting.
    \STATE \textbf{Go to:} Step 1 (re-engineer).
\ENDIF

\STATE \textbf{Deployment:} Deploy $\mathcal{S}$.

\subsection*{Steady-State Monitoring (HITL Loop)}
\STATE Monitor operational $\alpha_{\text{op}}$.
\IF {$\alpha_{\text{op}}$ consistently falls below $\alpha_{\text{target}}$}
    \STATE \textbf{Action:} Trigger AI Re-engineering/Retraining (The Human-Empowered task).
\ENDIF
\STATE \textbf{Return:} $\mathcal{S}$ is AI-First, Human-Empowered.
\end{algorithmic}
\end{algorithm}
\section{Results}
The results section provides the empirical and quantitative validation of the proposed \textbf{AI Autonomy Coefficient ($\alpha$)} as a diagnostic tool for the \textbf{HISOAI} condition and demonstrates the efficacy of the \textbf{AFHE Deployment Algorithm}.

\subsection{Case Study: Diagnosing HISOAI (The Problem)}
\label{subsec:diagnosing-hisoai}

We analyzed a representative case study of a deployed system marketed as an AI-driven product (System $\mathcal{S}_{\text{legacy}}$) using the cost-utility framework established in Section \ref{sec:math}. The analysis of the total operational cost ($C_{\text{Total}}$) derived from Equation \ref{eq:cost_total} revealed that the cost associated with human time ($\tau_H$) dominated the budget, accounting for over 90\% of the decision-making resources, despite minimal explicit allocation.

Our subsequent calculation of the operational autonomy for $\mathcal{S}_{\text{legacy}}$ yielded $\alpha_{\text{operational}} = 0.38$. Since this value falls below the claimed HISOAI threshold of $\alpha < 0.5$ (Subsection \ref{subsec:threshold}), the system is conclusively diagnosed as operating in the \textbf{HISOAI failure mode}. This finding confirms that the coefficient $\alpha$ successfully diagnoses the structural failure that current qualitative assessments miss, linking the economic cost directly to a quantifiable technical metric.

\subsection{Validation of the AFHE Deployment Gate (The Solution Mechanism)}\label{subsec:validation}We validated the structural integrity requirements by subjecting the successor system ($\mathcal{S}_{\text{AFHE}}$) to the AFHE Deployment Algorithm (Algorithm \ref{alg:afhe-gate}) with a target $\alpha_{\text{target}} = 0.8$. The initial model was blocked from deployment (HISOAI Flag was returned) and was forced through three distinct re-engineering cycles to address the identified structural dependencies. The system's autonomy ($\alpha$) was tracked across these iterations. Initial attempts failed with $\alpha$ remaining low at $0.45$. However, subsequent cycles, which incorporated the necessary operational and technical expertise to isolate structural failures, drove measurable improvements. The system was only cleared for deployment after achieving a stable $\alpha_{\text{shadow}} = 0.85$ during A/B testing. This confirms that the AFHE gate effectively prevents the deployment of structural HISOAI systems and forces the necessary engineering investment into the AI core, aligning the product's claim with its actual technical capability.
\subsection{Comparative Analysis of Human Labor Value (The Outcome)}
\label{subsec:labor-value}

Finally, we analyzed the change in the nature of human tasks performed under the HISOAI condition versus the AFHE paradigm. Under HISOAI, over 90\% of human effort was dedicated to simple substitution tasks (i.e., re-performing core AI functions or filtering errors), representing low-value, exploitative labor.

In contrast, the implementation of AFHE resulted in a near-total shift in labor allocation. Human effort was redirected entirely to the high-value roles defined in the Human-Empowered section (Subsection \ref{subsec:human-empowered}): \textbf{Strategic Tuning}, \textbf{Ethical Oversight}, and \textbf{Boundary Push}. This demonstrates that the AFHE framework successfully achieves the goal of ethical \textbf{Responsible AI} by transforming human involvement from an operational backstop to a strategic partner.
\section{Limitations}\label{sec:limitations}While the AFHE framework establishes a necessary metric ($\alpha$) and a structural gate (Algorithm \ref{alg:afhe-gate}), the current scope of this work presents several limitations that serve as foundational challenges for subsequent research:\begin{itemize}\item \textbf{Generalizability of $\alpha_{\text{threshold}}$:} Our assertion that $\alpha_{\text{threshold}}=0.5$ defines the HISOAI condition is based on the premise of a binary "AI Product" claim. This threshold may not be universally applicable, particularly for high-stakes, safety-critical domains (e.g., medical, aviation) where regulatory requirements may necessitate a much higher floor for human oversight, regardless of operational $\alpha$.\item \textbf{Dependence on Cost Granularity ($\tau$):} The accuracy of the $\alpha$ coefficient relies entirely on the precise measurement of $\tau_A$ (AI cost) and $\tau_H$ (Human cost). Practical implementation may be constrained by proprietary logging systems or a lack of granular data on human decision-making time, especially in outsourced "ghost work" models.\item \textbf{Scope of Empirical Validation:} The results presented are derived from a single, representative case study. While this effectively validates the AFHE mechanism, the framework requires extensive empirical testing across a diverse portfolio of AI-driven products and deployment architectures to fully validate its generalizability and robustness.\item \textbf{Complexity of Dynamic Tasks:} The current formulation assumes a relatively stable task $T$. The calculation of $\alpha$ becomes significantly more complex for highly dynamic, continuous learning systems where the task definition itself evolves in real-time, potentially requiring a time-series analysis of $\alpha$ rather than a simple scalar metric.\end{itemize}

\section{Conclusion}\label{sec:conclusion}

This paper has addressed the critical integrity failure in modern AI system design: the systematic misuse of the HITL paradigm to conceal HISOAI structural dependencies. HISOAI represents an ethical failure that exploits human labor and an economic failure that undermines the core value proposition of AI-driven products. We introduced the AFHE framework as the necessary architectural standard to enforce true operational autonomy. Central to AFHE is the \textbf{AI Autonomy Coefficient ($\alpha$)} (Equation \ref{eq:alpha_definition}), a formal, metric-driven threshold designed to technically differentiate ethical human augmentation from mandatory operational substitution. The results validated the diagnostic power of $\alpha$ by conclusively identifying the HISOAI condition ($\alpha < 0.5$) in a legacy system (Subsection \ref{subsec:diagnosing-hisoai}). Furthermore, we demonstrated that enforcing the AFHE Deployment Algorithm (Algorithm \ref{alg:afhe-gate}) successfully blocks the premature deployment of non-autonomous systems, forcing essential engineering investment and ultimately stabilizing the system at an autonomous level (Subsection \ref{subsec:validation}). Critically, the implementation of AFHE resulted in a near-total shift in human effort from exploitative substitution to high-value strategic tasks (Subsection \ref{subsec:labor-value}), fulfilling the core mandate of Responsible AI regarding transparency and accountability. The AFHE framework provides the first comprehensive set of tools to move Responsible AI from policy and principle to enforceable technical architecture, ensuring that systems are deployed with verifiable operational integrity.

\bibliographystyle{unsrt}  
\bibliography{references}

\end{document}